\documentclass[a4paper]{article}

\usepackage[utf8]{inputenc}
\usepackage{kotex}
\usepackage{booktabs}
\usepackage{caption}
\usepackage{subcaption}
\usepackage{multirow}
\usepackage[hang]{footmisc}
\usepackage[noadjust]{cite}
\usepackage{hyperref} % URL in footnote\
\usepackage{algorithm,algpseudocode}% 

\usepackage{INTERSPEECH2022}

\title{Enhancement of Pitch Controllability \\using Timbre-Preserving Pitch Augmentation in FastPitch}
\name{Hanbin Bae, Young-Sun Joo}
\address{
  Speech AI Lab, NCSOFT Corp., Republic of Korea}
\email{bhb0722@ncsoft.com, ysjoo555@ncsoft.com}

\begin{document}

 \maketitle
\begin{abstract} 
The recently developed pitch-controllable text-to-speech (TTS) model, i.e. FastPitch, was conditioned for the pitch contours. However, the quality of the synthesized speech degraded considerably for pitch values that deviated significantly from the average pitch; i.e. the ability to control pitch was limited. 
To address this issue, we propose two algorithms to improve the robustness of FastPitch. 
First, we propose a novel timbre-preserving pitch-shifting algorithm for natural pitch augmentation. Pitch-shifted speech samples sound more natural when using the proposed algorithm because the speaker's vocal timbre is maintained. 
Moreover, we propose a training algorithm that defines FastPitch using pitch-augmented speech datasets with different pitch ranges for the same sentence. 
The experimental results demonstrate that the proposed algorithms improve the pitch controllability of FastPitch. 
\end{abstract}


\section{Introduction}
\label{sec_introduction}

Dynamic pitch contour is a crucial factor in synthesizing natural speech. Therefore, recently, studies have been actively conducted on pitch-controllable text-to-speech (TTS) models have been actively researched \cite{fastspeech2,fastpitch,tjpark}. Among them, FastPitch \cite{fastpitch}, that controls the phoneme-level pitch and duration of synthesized speech by conditioning the pitch and duration values, has recently become popular.

However, in our pilot experiments, we observed that larger pitch adjustment values in FastPitch significantly degraded the quality of the synthesized speech. For example, the pronunciation clarity and intelligibility of the synthesized speech were reduced when the input pitch values deviated considerably from the average pitch range; further details on these results are discussed in Section~\ref{sec_experiment}. 
This problem is primarily caused by the unbalanced ratio of the speech database. 
Generally, speech samples in a database are mainly distributed around an average pitch value, and the number of speech samples with pitch ranges that are significantly different from the average value are relatively small. This distribution prevents FastPitch from learning sufficiently over a wide pitch range; the model's ability to control pitch is thus limited.

A simple approach to address this issue is to record a sufficient number of speech samples for various pitch ranges. However, building such a database is not favorable under realistic conditions, owing to the cost and time required. 
Data augmentation can be an alternative approach.
In \cite{hsu2019disentangling, valentini2016speech,adiga2019speech,bgmtts}, 
the speech data obtained from external sources such as audiobooks and crowd-sourced data was utilized to train TTS models. These studies primarily focused on increasing the quantity of speech database but enhancing the pitch controllability of FastPitch is necessary to augment speech data over a wide range of pitches. The pitch augmentation can be achieved by shifting the pitch of speech data.  The key point here is that the pitch-shifted speech should be perceptually similar to the speaker's vocal timbre, which is known as vocal color.
In this paper, these pitch-shifting (PS) algorithm is called a timbre-preserving PS algorithm. 

Timbre-preserving PS algorithms are classified into two main types: parametric and non-parametric methods. First, parametric methods are based on parametric vocoders, such as WORLD \cite{world} vocoder. These method extract vocoder parameters, also called acoustic features, from the speech signal, and then, re-synthesize the pitch-shifted speech with parameters including modified pitch values. Second, non-parametric methods are performed directly on time-domain signals. The time-domain pitch-synchronous overlap and add (TD-PSOLA) technique \cite{tdpsola} is utilized. The TD-PSOLA technique divides pitch-synchronous windowed speech segments at pitch marks, which are time stamps placed at the peaks of the periodic signals, and then, combines the speech segments at the pitch marks adjusted to the desired pitch interval.

These methods successfully control the pitch of the speech signals, while preserving the vocal timbre. Despite this successful pitch-control ability, the quality of the resultant pitch-shifted speech generated by these methods depends on the vocoder models or on the degree of pitch adjustment. In the parametric method, regardless of the PS, a well-designed vocoder model is required to obtain synthesized speech with high audible quality. 
In addition, even with a well-designed vocoder model, the quality of generated speech is affected by factors including inaccurate vocoded parameter values estimated using pitch or spectral envelope estimation algorithms. In the TD-PSOLA method, a slight pitch adjustment maintains the speaker's timbre; however, an excessive pitch adjustment affects the timbre of the pitch-shifted speech.

To address these issues, we propose (1) a novel timbre-preserving PS algorithm that does not require extra algorithms, such as pitch tracking or spectral envelope estimation, and (2) a FastPitch training algorithm that uses the pitch-augmented speech datasets. Based on our proposed methods, FastPitch can generate a high-quality synthesized speech even for larger pitch adjustment values. In other words, it enhances the pitch controllability of FastPitch. 
More detail processes are as follows. First, the pitch-augmented speech datasets are obtained using the proposed PS algorithm, which utilizes the characteristics of the VocGAN \cite{vocgan}, a neural vocoder. 
This algorithm raises or lowers the pitch of the speech samples, while maintaining the speaker's vocal timbre. Since the proposed algorithm is not based on a parametric vocoder, it has the advantage of being less affected by inaccurate by inaccurately extracted vocoder features. Also, the pitch adjustment range is relatively less restricted compared to the TD-PSOA method. 
After augmenting the speech data for a wide range of pitches, FastPitch is trained using a training algorithm that was proposed specifically for pitch-augmented datasets. FastPitch is updated using the original and pitch-augmented datasets alternately for every epoch. Because the pitch-augmented datasets comprise speech samples with different pitches for the same sentence, the pitch and duration predictors are fixed while updating the FastPitch with them. Please note that the duration and pitch predictors cannot predict different values for the same text.
\section{Proposed Method}
\label{sec_NeurFPPS}

\subsection{VocGAN-based PS algorithm}
\label{sec_vocgan-ps}

\begin{figure}[tb]
    \centering
    \centerline{\includegraphics[width=7cm]{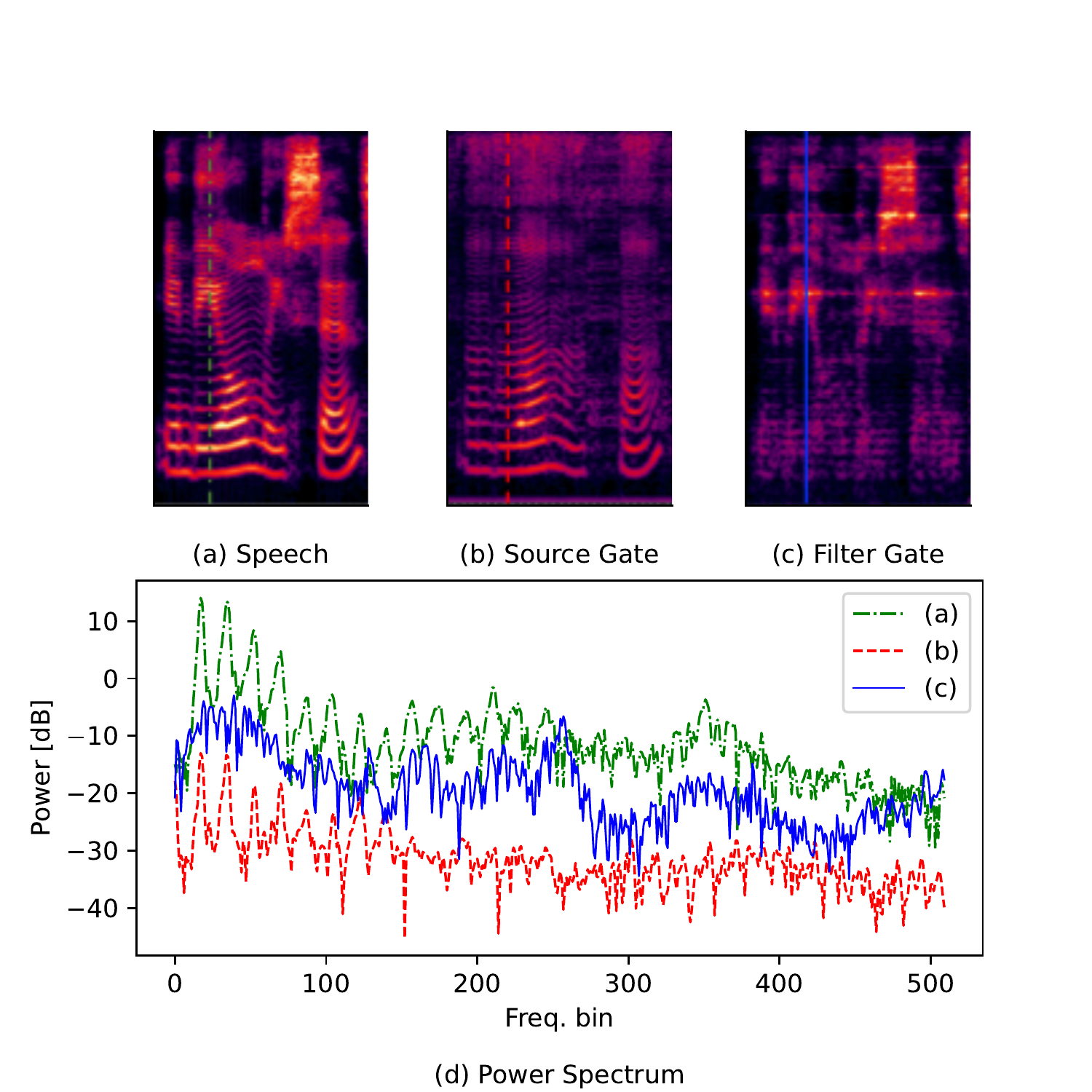}}
\caption{Mel-spectrograms, where (a) represents recorded speech, (b) and (c) represent the output waveforms from the source and filter gates, respectively, and (d) represents their power spectra.}
\label{fig_mel-gate-output}
\end{figure}

VocGAN \cite{vocgan} is a neural vocoder comprising a multi-scale waveform generator and discriminator. 
In the model training step, five waveforms that are generated with low to high resolutions are used for training. In the inference step, only the waveform with the highest resolution is used.
In our preliminary experiments, 
we referred the two low-resolution gates as the source gate; and the three high-resolution gates as the (vocal tract) filter gate.
We applied the same mel-spectrogram to the source and filter gates of the pre-trained VocGAN model and perceptually observed that each waveform had different characteristics.
The output signal from the source gate had a pitch that was perceptually similar to the input speech; however, it contained less linguistic information. In contrast, the output signal from the filter gate contained more linguistic information from the input speech without a pitch-related information. 
Figure~\ref{fig_mel-gate-output} depicts the mel-spectrograms of the input speech and the output signals of the source and filter gates.

\begin{figure}[tb]
     \centering
     \begin{subfigure}[tb]{0.3\textwidth}
         \centering
         \centerline{\includegraphics[width=7cm]{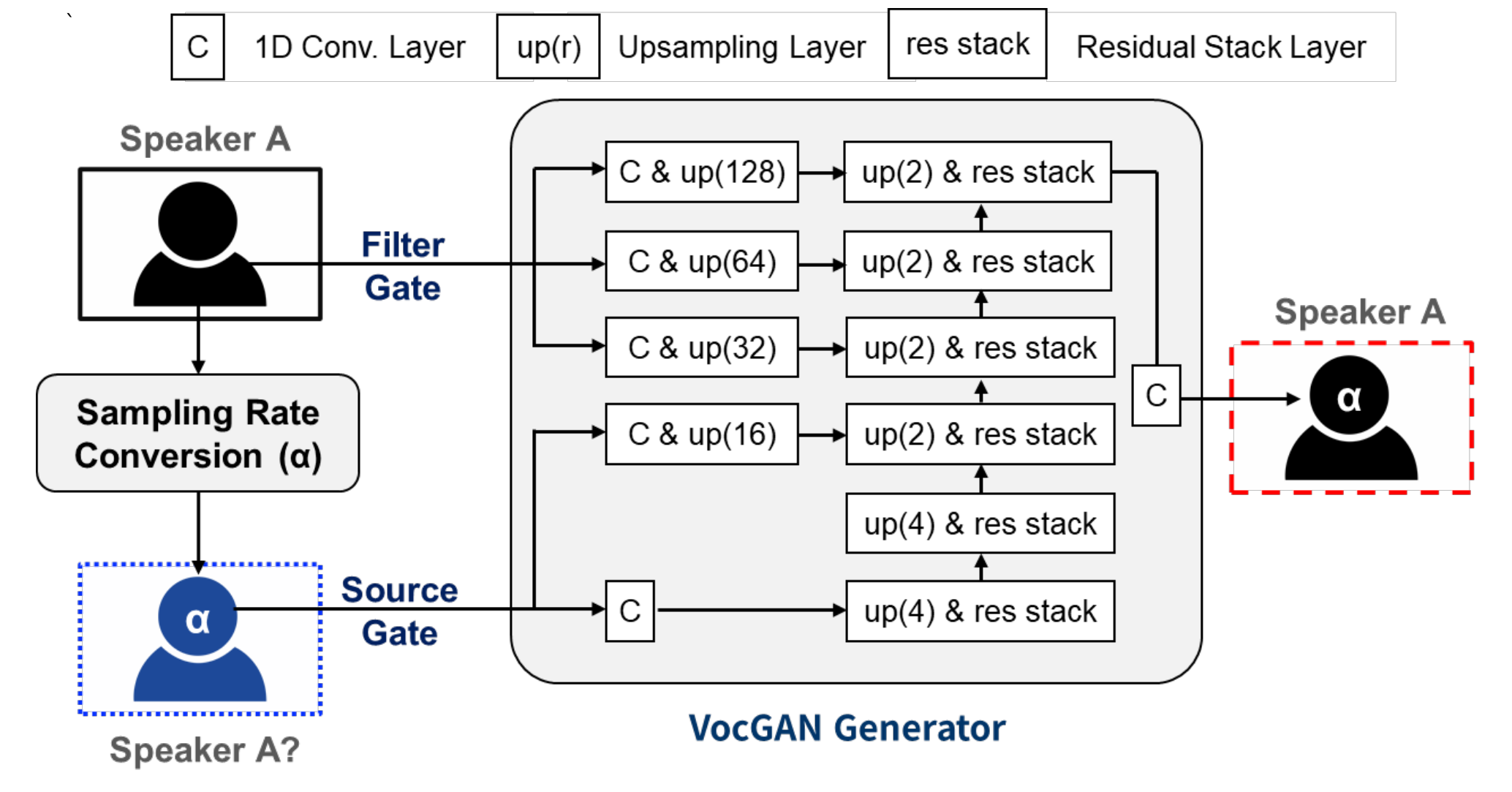}}
     \end{subfigure}
     \hfill
     \begin{subfigure}[tb]{0.3\textwidth}
         \centering
         \centerline{\includegraphics[width=7cm]{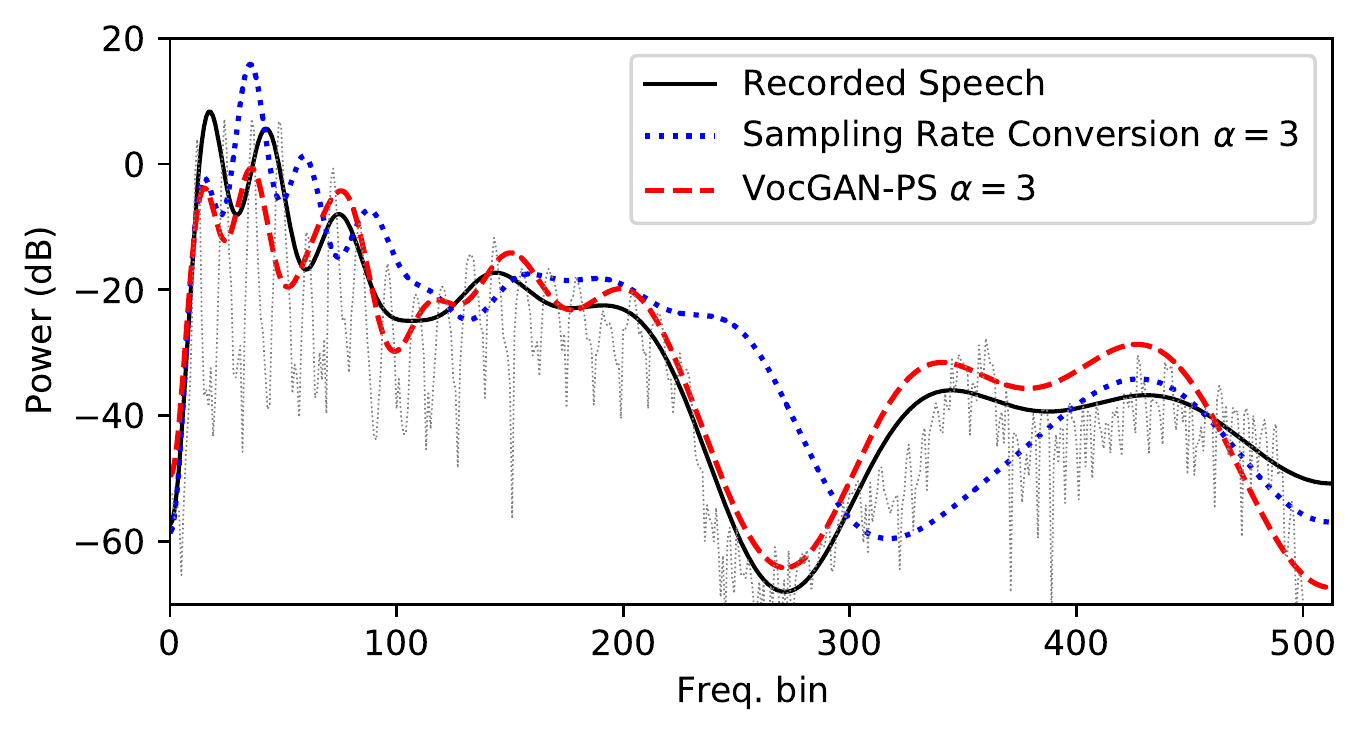}}
     \end{subfigure}
         \caption{Schematic of the VocGAN-PS and Spectral envelopes of the three types of speech waveforms. }
    \label{fig_fpps-algorithm}
\end{figure}

Based on these characteristics, we propose the VocGAN-based PS algorithm (VocGAN-PS), which shifts the pitch of the input speech according to the pitch adjustment value $\alpha$, while maintaining the speaker's timbre by preserving the spectral envelopes; the $\alpha$ is represented in semitone (ST) units.
Figure~\ref{fig_fpps-algorithm} depicts the schematic of the VocGAN-PS.
For the filter gate, the original mel-spectrogram is applied to preserve the spectral envelopes, whereas for the source gate, the pitch-shifted mel-spectrogram is applied to obtain the target pitch information. The pitch-shifted mel-spectrogram is obtained by the sampling rate conversion \cite{sr-conversion} in the frequency-domain.
Finally, the output speech of VocGAN is obtained with the correct timbre of the speaker.

\subsection{Availability of pitch-augmented data for training}
\label{sec_availability}

\begin{figure*}[t]
    \centering
    \centerline{\includegraphics[width=15cm]{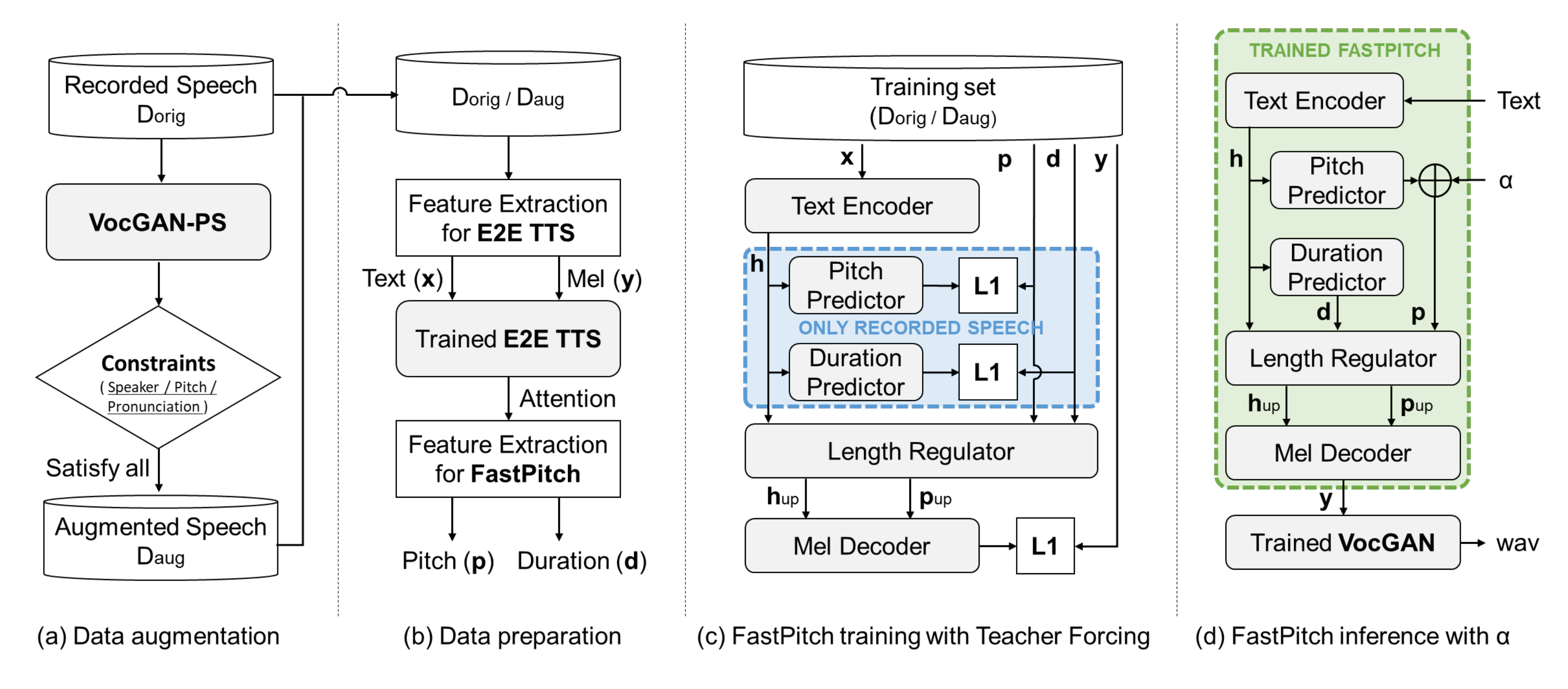}}
\caption{Schematic diagram of proposed method.}
\label{fig_diagram}
\end{figure*}
Although the VocGAN-PS preserves the vocal timbre of the input speech well, the training data of the TTS model should have high pronunciation accuracy and should be perceptually similar to the target speaker's speech. Therefore, we evaluate the speech quality and speaker similarity for the pitch-shifted speech samples and use only those samples that meet the constraints as the augmented dataset, as shown in Figure~\ref{fig_diagram}a.

\subsection{Data preparation}
Figure~\ref{fig_diagram}b depicts the process of the data preparation. First, we extracted the character sequence $\mathbf{x}$ and mel-spectrogram $\mathbf{y}$. The mel-spectrogram using an 80-bins mel filter bank, which was computed by the FFT with size of $1024$, hop size of $256$, and window size of $1024$ in the frequency range $0-8000$ Hz. We then extracted the character-level durations $\mathbf{d}$ and pitches $\mathbf{p}$ using the attention matrix of the pre-trained deep-convolutional TTS model \cite{dctts}, which can easily train an attention matrix by guided-attention loss.

\subsection{Training FastPitch using pitch-augmented dataset}

Figure~\ref{fig_diagram}c depicts the proposed training algorithm. 
The FastPitch $\mathcal{M}_{\mathrm{FastPitch}}$ is updated using the original and augmented datasets, namely $D_{\textrm{orig}}$ and $D_{\textrm{aug}}$, alternately for 1 epoch each. 

For $D_{\textrm{orig}}$, the text encoder first outputs a hidden representation $\mathbf{h}$ corresponding to the character sequence $\mathbf{x}$ of the input text. The pitch predictor and duration predictor then predict the duration sequence $\hat{\mathbf{d}}$ and pitch sequence $\hat{\mathbf{p}}$, respectively, corresponding to the input $\mathbf{h}$. For training, the teacher forcing technique \cite{teacherforcing} was used; this means that the ground-truths $\mathbf{d}$ and $\mathbf{p}$ were used for upsampling and as the inputs to the mel decoder. Then, $\mathbf{h}_{\mathrm{up}}$ and $\mathbf{p}_{\mathrm{up}}$, which are upsampled using a length regulator, pass through the mel decoder that predicts the output mel-spectrogram $\hat{\mathbf{y}}$.

For $D_{\textrm{aug}}$, $\mathcal{M}_{\mathrm{FastPitch}}$ generates $\mathbf{\hat{y}}$ by applying the pitch-shifted pitch $\mathbf{p}+\alpha$. In this step, the losses are calculated for the pitch-shifted mel-spectrograms only. Note that the duration and pitch predictors are not used in this step because it is impossible for the pitch predictor to predict $\mathbf{p}+\alpha$ instead of $\mathbf{p}$ for the same $\mathbf{h}$.
\section{Experiments} \label{sec_experiment}

\begin{figure}[tbp]
    \centering
     \centerline{\includegraphics[width=7cm]{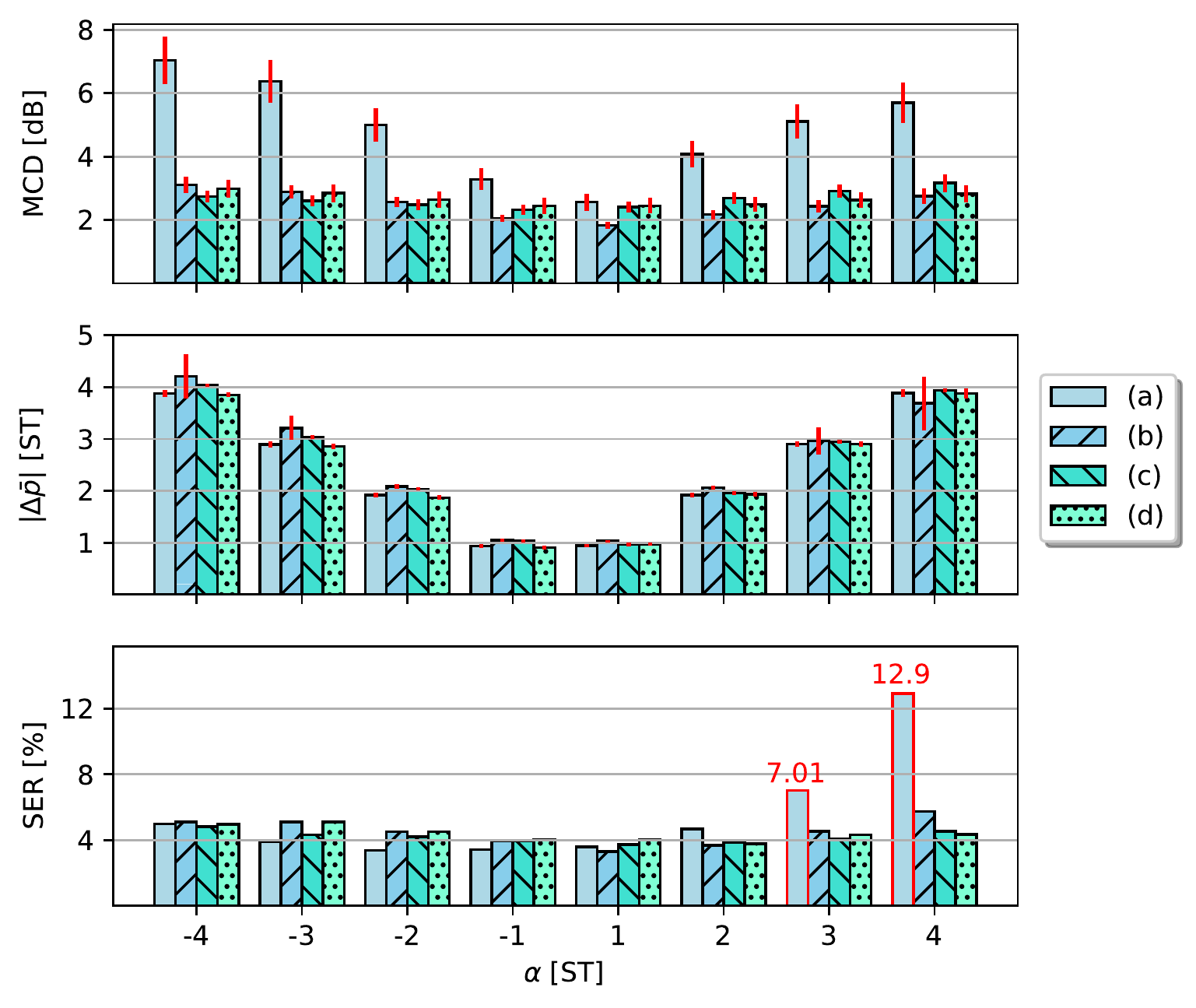}}
\caption{Objective evaluation results for the PS algorithms, where (a), (b), (c) and (d) represent the SoX-, TD-PSOLA-, WORLD- and VocGAN-PSs, respectively.}
\label{fig_mcd_ser}
\end{figure}

In this study, we used $14000$ speech samples of a single Korean female speaker (target speaker), which were sampled at $22.05$ kHz, amounting to approximately $20$ h. The dataset was divided into the training, validation, and test sets with $13400$, $100$, and $500$ samples, respectively.
The distribution of mean-referred pitch values of target speaker is $\mathbf{p} \sim \mathcal{N}(0, 4.11^{2})$, which is represented in ST units by the equation $\mathbf{p} = 12\cdot\mathrm{log}_{2}(\mathbf{f}_{0}/\bar{\mathbf{f}}_{0})$, with the average fundamental frequency $\bar{\mathbf{f}}_{0}$ = $248$ Hz.

\subsection{Pitch augmentation using the VocGAN-PS}
\label{sec_vocganps_performance}

Before generating the augmented dataset for the training set, we generated the pitch-shifted speech samples using $500$ samples of the test set for $|\alpha| \leq 4$ (less than $1$ standard deviation).
We then conducted the objective and subjective tests to evaluate the speech quality and speaker similarity of the pitch-shifted speech samples as mentioned in Section~\ref{sec_availability}.

The SoX- (vocal timbre is not preserved), TD-PSOLA- and WORLD-PS algorithms were also evaluated to compare with the VocGAN-PS. We used the open sources~\cite{sox, os_tdpsola, os_world} and its default settings except for the VocGAN-PS. In case of the VocGAN, we implemented it and trained using the same training set.

\begin{table}[t]
  \caption{MOS ($\uparrow$) results for speech quality of the three timbre-preserving PS algorithms with 95\% confidence intervals (CIs).}
  \label{table_ps-mos}
  \centering
  \begin{tabular}{cccccccc}
  \toprule
  \multicolumn{1}{c}{\textbf{}} &
  \multicolumn{1}{c}{\textbf{$\mathbf{-4}$}} &
  \multicolumn{1}{c}{\textbf{$\mathbf{-3}$}} &
  \multicolumn{1}{c}{\textbf{$\mathbf{-2}$}} &
  \multicolumn{1}{c}{\textbf{$\mathbf{+2}$}} & 
  \multicolumn{1}{c}{\textbf{$\mathbf{+3}$}} & 
  \multicolumn{1}{c}{\textbf{$\mathbf{+4}$}} \\
  \midrule
 
 \multirow{2}{*}{{\shortstack[c]{TD-PSOLA\\$\pm$ CI} }} & 
  $1.49$ &	$1.69$ &	$1.93$ &	$1.99$ &	$1.67$ & $1.73$ \\
& $0.13$ & $0.12$ & $0.14$ & $0.16$ & $0.13$  & $0.14$ \\
  \midrule 
  
  \multirow{2}{*}{{\shortstack[c]{WORLD\\$\pm$ CI} }} & 
  $2.64$ &	$3.16$ &	$2.95$ &	$3.05$ &	$3.11$ & $2.81$ \\
& $0.17$ & $0.18$ & $0.16$ & $0.17$ & $0.18$  & $0.17$  \\
  \midrule 
  
  \multirow{2}{*}{{\shortstack[c]{VocGAN\\$\pm$ CI} }} & 
$\mathbf{3.61}$ &	$\mathbf{4.03}$ &	$\mathbf{4.35}$ &	$\mathbf{4.35}$ &	$\mathbf{4.11}$ & $\mathbf{3.85}$ \\
& $0.15$ & $0.13$ & $0.14$ & $0.13$ & $0.15$  & $0.15$  \\

  \bottomrule
  \end{tabular}
\end{table}
\begin{figure}[tb]
    \centering
    \centerline{\includegraphics[width=7cm]{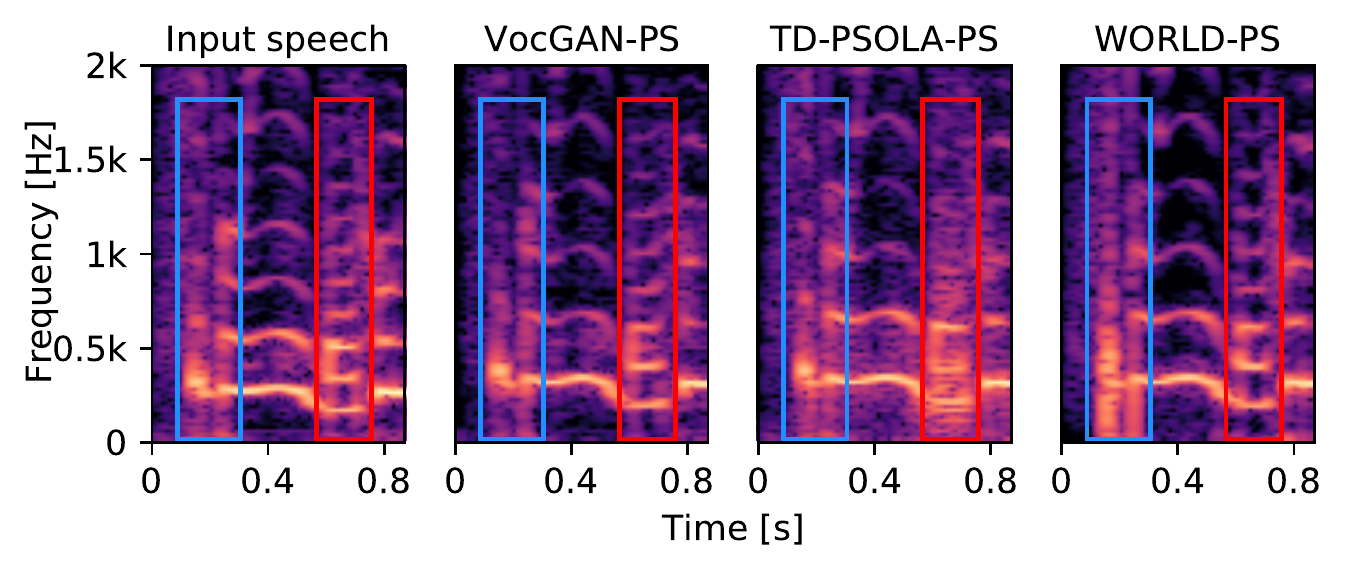}}
\caption{Spectrogram of the input speech and the pitch-shifted samples of each algorithm, where $\alpha=3$. }
\label{fig_tpps_specs}
\end{figure}
\begin{figure}[t]
    \centering
    \centerline{\includegraphics[width=7cm]{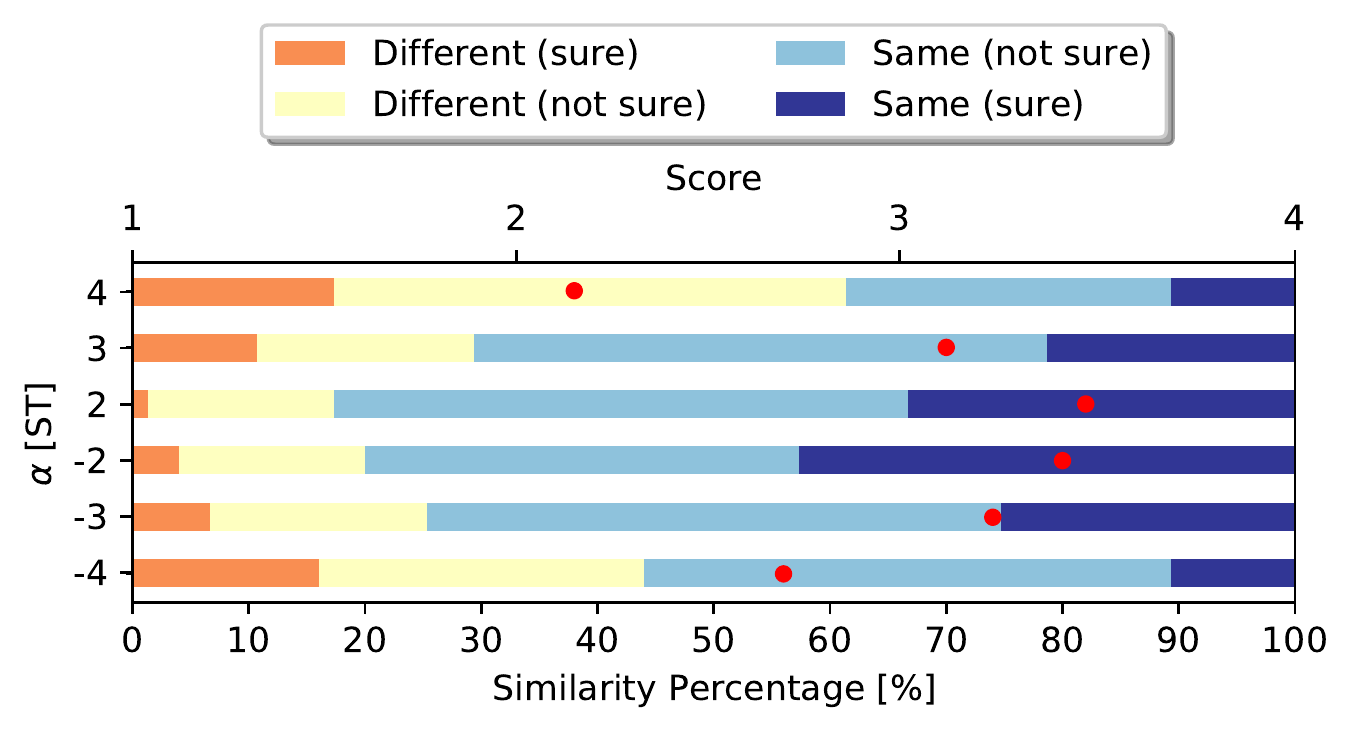}}
\caption{Similarity results of the VocGAN-PS. Red dots represent their mean value.}
\label{fig_sim}
\end{figure}
\begin{figure}[t]
    \centering
    \centerline{\includegraphics[width=7cm]{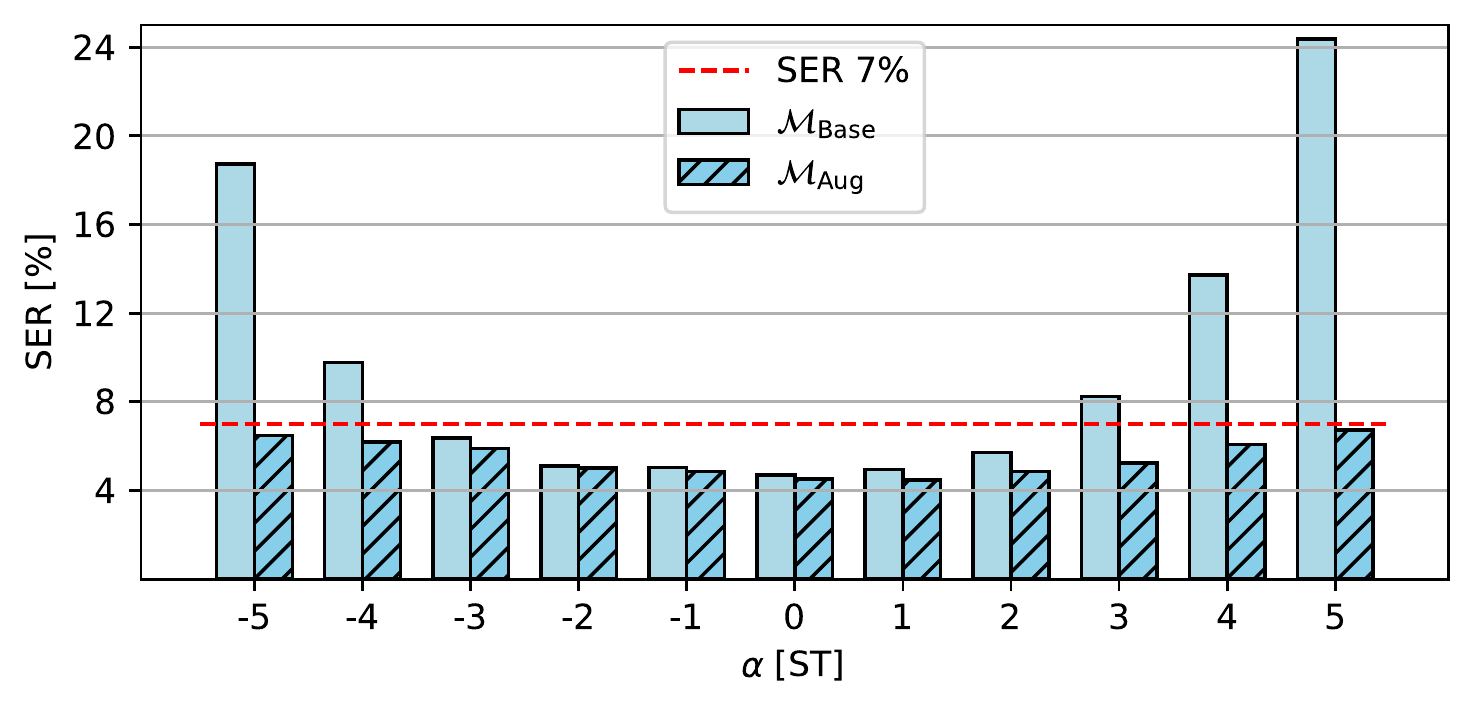}}
\caption{SER results for pronunciation accuracy.}
\label{fig_ser}
\end{figure}

To evaluate the speech quality, we first conducted the objective test calculating the mel-cepstral distortion (MCD)~\cite{kubichek1993mcd}, change in the average pitch ($\Delta\bar{\mathbf{p}}$), and syllable error rate (SER) using the Kaldi speech recognition system \cite{kaldi}.
Second, we conducted the mean opinion score (MOS) test on the speech quality by randomly selecting five samples from the test set for the three timbre-preserving PS algorithms of $\alpha \in \{\pm2, \pm3, \pm4\}$ for a total of $30$ samples for each algorithm. A total of 15 native Koreans participated in this effort and scored the samples on a scale of $1$ (Bad) to $5$ (Excellent) referred by \cite{crowdmos}.

These results are summarized in Figure~\ref{fig_mcd_ser} and Table~\ref{table_ps-mos}. Comparing the timbre-preserving algorithms for all $\alpha$, the speech quality of the VocGAN-PS outperform the other PS algorithms overall.
Figure~\ref{fig_tpps_specs} depicts the spectrogram of the input speech and pitch-shifted samples of each algorithm.
Comparing the areas in the red box, where the pitch values were lower than about $200$ Hz, the TD-PSOLA-PS generated the low speech quality samples caused by the PSOLA technique.
Comparing the areas in the blue box, where is the unvoiced/voiced transition, the WORLD-PS occurred the distortion caused by the inaccurately estimated pitch values.

To evaluate the speaker similarity, we conducted the listening test to evaluate the speaker similarity between the input recorded speech and pitch-shifted speech. We selected five samples for every value of $\alpha \in \{\pm2, \pm3, \pm4\}$ randomly for a total of $30$ samples.
A total of 15 native Koreans participated in this effort and scored the samples on a four-point scale: 1 (Different, absolutely sure), 2 (Different, not sure), 3 (Same, not sure) and 4 (Same, absolutely sure) referred by \cite{vcc2020}; these results are summarized in Figure~\ref{fig_sim}. For $|\alpha|\leq3$, 
about $70\%$ or more of the pitch-shifted samples were judged by participants to be the same as the target speaker.

Hence, in subsequent experiments, we used the pitch-shifted speech samples for $|\alpha| \leq 3$ as the augmented dataset.

\subsection{Performance evaluation of the FastPitch}
\label{sec_eval_fastpitch}

\begin{table}[t]
  \caption{MOS results for speech quality of the two FastPitch models with 95\% CIs.}
  \label{table_mos-speech-quality}
  \centering
  \begin{tabular}{cccccccc}
  \toprule
  \multicolumn{1}{c}{\textbf{$\alpha$}} &
  \multicolumn{1}{c}{\textbf{$\mathbf{-4}$}} &
  \multicolumn{1}{c}{\textbf{$\mathbf{-2}$}} &
  \multicolumn{1}{c}{\textbf{$\mathbf{0}$}} &
  \multicolumn{1}{c}{\textbf{$\mathbf{+2}$}} & 
  \multicolumn{1}{c}{\textbf{$\mathbf{+4}$}} \\
  \midrule
  \multirow{2}{*}{{\shortstack[l]{$\mathcal{M}_{\mathrm{Base}}$\\ \hspace{0.5cm} $\pm$ CI} }} & 
  $2.88$ &	$3.97$ &	$4.46$ &	$4.05$ &	$3.18$ \\
              
& $0.18$ & $0.15$ & $0.16$ & $0.16$ & $0.18$  \\
  \midrule 
 
  \multirow{2}{*}{{\shortstack[l]{$\mathcal{M}_{\mathrm{Aug}}$\\ \hspace{0.5cm} $\pm$ CI} }}  &  $\mathbf{3.63}$ &	$\mathbf{4.27}$ &	$\mathbf{4.58}$ &	$\mathbf{4.15}$ &	$\mathbf{3.55}$ \\
          & $0.16$ & $0.15$ & $0.14$ & $0.14$ & $0.17$  \\
  \bottomrule
  \end{tabular}
\end{table}

We compared $\mathcal{M}_{\mathrm{Base}}$ and $\mathcal{M}_{\mathrm{Aug}}$ trained by the basic FastPitch algorithm and proposed training algorithm, respectively. Both models were trained up to $300$ epochs each with a mini-batch size of $16$ and the Adam optimizer \cite{adam} with initial learning rate of $0.0002$. The hyper-parameters of Adam optimizer were $({\beta}_1,{\beta}_2)=(0.5, 0.9)$, and ${\epsilon}=10^{-6}$.

To evaluate the pitch controllability of FastPitch, 
we explored the maximum pitch adjustment value $\alpha^{*}$ for each model. This means that the model generates speech for $\hat{\mathbf{p}}\pm\alpha^{*}$ without performance deterioration of pronunciation clarity and speech quality. We set a range in which (1) the SER is lower than $7\%$ and (2) the speech quality score is higher than $3$ (Fair) as $\alpha^{*}$. 
We observed that the pitch-shifted speech generated by the SoX-PS algorithm had SERs greather than $7$\% without the deterioration of pronunciation clarity; however, its timbre sounded perceptually inhuman or unnatural similar to the effect of helium on a human voice. 
Therefore, we set an SER of $7\%$ as the constraint for evaluating the pronunciation clarity of the FastPitch models.

To explore $\alpha^{*}$, we generated speech samples for $500$ texts of the test set by adding $\alpha$ to $\mathbf{\hat{p}}$, as shown in Figure~\ref{fig_diagram}d; $\alpha$ was set from $-5$ to $+5$. 
Figure~\ref{fig_ser} depicts the SERs of both models. For $\mathcal{M}_{\mathrm{Aug}}$, the SER is maintained lower than $7\%$ for all $\alpha$, whereas for $\mathcal{M}_{\mathrm{Base}}$, the SER increases as $\alpha$ increases.

We conducted the MOS test on speech quality by randomly selecting eight samples for each case of $\alpha \in \{0, \pm2, \pm4\}$ for a total of $40$ samples for each model; this test is the same participant and scale in Section~\ref{sec_vocganps_performance}.
Table~\ref{table_mos-speech-quality} shows that the results of the MOS test.
$\mathcal{M}_{\mathrm{Aug}}$ model outperforms the $\mathcal{M}_{\mathrm{Base}}$ model overall; in addition, $\mathcal{M}_{\mathrm{Aug}}$ received a score of $3.63$ even when $\alpha$ was equal to $-4$, whereas the quality of $\mathcal{M}_{\mathrm{Base}}$ was significantly degraded.

In summary, the value of $\alpha^{*}$ for $\mathcal{M}_{\mathrm{Aug}}$ was determined as $4$ and that for $\mathcal{M}_{\mathrm{Base}}$ was $2$; this proved that the proposed training algorithms improved the pitch controllability of FastPitch. 

Audio samples can be found online\footnote{\url{https://nc-ai.github.io/speech/publications/vocgan-ps-fastpitch}}.
\section{Conclusion}
\label{sec_conclusion}
This paper proposed two algorithms to enhance the robustness of pitch controllability of FastPitch. The proposed VocGAN-PS first augmented the speech data successfully such that it contained a wide range of pitches while preserving vocal timbre. Then, a training algorithm was proposed for FastPitch such that the pitch-augmented data could be used.
Thus, FastPitch generated speech with stable quality even at higher and lower pitch ranges.
We verified the effectiveness of the proposed methods through various quantitative and qualitative evaluations.

\bibliographystyle{IEEEtran}
\bibliography{refs}

\end{document}